 \documentstyle[prd,eqsecnum,preprint,tighten,aps]{revtex}
 
\begin{document}
 \renewcommand{\theequation}{\thesection.\arabic{equation}}
 \def\appendixa{
 \vskip 1cm
 {\bf APPENDIX A: POINCAR\'E BI-ALGEBRA FROM $h(1)$} \vskip 1cm
 \par
 \setcounter{equation}{0}
 \def\theequation{A.\arabic{equation}}
 }
 \def\appendixb{
 \vskip 1cm
 {\bf APPENDIX B: PROPERTIES OF THE SET OF FUNCTIONS $\{F(k, p)\}$}
 \par
 \setcounter{equation}{0}
 \def\theequation{B.\arabic{equation}}
 }

  \title{Quantization of Scalar Fields in Curved Background and
 Quantum Algebras}
 \author{A. Iorio$^{a,b,}$\thanks{E-mail: iorio@sa.infn.it, iorio@maths.tcd.ie},
 G. Lambiase$^{a,b,}$\thanks{E-mail: lambiase@sa.infn.it}, and
 G.Vitiello$^{a,b,c,}$\thanks{E-mail: vitiello@sa.infn.it} }
  \address{$^a$Dipartimento di Fisica "E.R. Caianiello"
  Universit\`a di Salerno, 84081 Baronissi (Sa), Italy.}
  \address{$^b$INFN, Gruppo Collegato di Salerno, and $^c$INFM, Salerno, Italy.}
 \date{\today}
 \maketitle
 \begin{abstract}
 We show that a suitable deformation of the algebra $h_k(1)$ of the
 creation and annihilation operators for a complex scalar field,
 initially quantized in Minkowski space--time, induces the
 canonical quantization of the same field in a generic
 gravitational background. This discloses the physical meaning of the
 deformation parameter $q$ which turns out to be related to the
 gravitational field. The thermal properties are re-obtained in
 this formalism, and the application to Schwarzschild and Rindler
 space-times are carried out.
 \end{abstract}

 \section{Introduction}
 \setcounter{equation}{0}

Although many attempts have been made to quantize gravity, a
satisfactory and definitive theory still does not exist. As well
known indeed, one of the most discussed problems is the
non-renormalizability of General Relativity when quantized as a
local Quantum Field Theory (QFT).

In the absence of a theory of quantum gravity, one can try to
construct an effective theory in which higher--order terms of
curvature invariants such as $R^2, R^{\mu\nu}R_{\mu\nu},
R^{\mu\nu\alpha\beta}R_{\mu\nu\alpha\beta}, R\Box R$, $R\Box^kR$,
or non-minimally coupled terms between scalar fields and geometry,
as $\varphi R$, $\varphi^2R$, appear in the action
\cite{brans,vilkovisky,stelle,christensen,barth,odintsov,capozziello}.
Any action involving such terms generalizes the Einstein--Hilbert
action of General Relativity, which is linear in the Ricci scalar
$R$, and represents a low--energy approximation to some
fundamental theory of gravity. For example, string theory or
supergravity present low--energy effective actions where
higher--order or non-minimally coupled terms appear
\cite{fradkin}.

On the other hand, quantum aspects of gravity can be investigated
by studying QFT in curved space-time. In this {\it semiclassical}
approach, one analyzes the quantization of matter fields in the
presence of the gravitational field, which is treated as a
classical background described by General Relativity
\cite{birrell}. One of the most important consequences is that
quantum effects lead to thermal evaporation of black holes
\cite{HAW}, with a temperature (the Bekenstein-Hawking
temperature) given by, in units $\hbar=c=k_B=1$,
\begin{equation}\label{1a}
T_{BH}=\frac{k}{2\pi}\,{,}
\end{equation}
where $k$ is the surface of gravity of a black hole ($k \sim
(GM)^{-1}$). This effect was soon realized to be associated with
the existence of an event horizon in Schwarzschild space--time.
Owing to this result, different background space-times have been
investigated, especially the Rindler space-time, that is a flat
space--time with an horizon, associated with a uniformly
accelerated observer in Minkowski space-time. Davies \cite{DAV}
and Unruh \cite{UNR} have shown that the vacuum state for an
inertial observer is a canonical ensemble for an uniformly
accelerated (Rindler) observer. The temperature  $T_R$
characterizing  this ensemble is related to the acceleration of
the observer by the relation
\begin{equation}\label{1}
T_R\,=\,\frac{a}{2\pi} \;.
\end{equation}
This is the thermalization theorem, in a nutshell (for a review
see \cite{TAK}).

The purpose of this paper is to show that, in the framework of the
semiclassical theory of gravity, quantum algebras are suitable
structures to quantize a scalar field in presence of a
gravitational background. They are important for several reasons.
Firstly, they induce event horizons in the space--time. Secondly,
a deformation of the canonical algebra of a quantized complex
scalar field, in flat space--time, induces, in a surprisingly
natural way, the canonical quantization of the same field in a
generic curved space-time. Finally, the deformation parameter $q$
can be related to the gravitational field. Thus thermal properties
of QFT in curved space--time can be derived in this new setting.

In recent years there has been a deep investigation of the role
played by deformed algebraic structures in General Relativity, see
for instance Ref. \cite{WESS,MUS}. Wess and Cerchiai \cite{WESS}
analyze the non-commutative structure of the Minkowski space-time
emerging from quantum group considerations. In that approach the
Lorentz group is $q$-deformed and the corresponding space-time
variables (coordinates and momenta) satisfy the commutation
relations of a non-commutative space {\it \`a la} Connes
\cite{connes}. On the other hand, Musto et al. \cite{MUS} consider
the $q$-deformation of the Poincar\'e group of non-commuting
tetrads. They show that General Relativity appears as a common,
invariant sector of a one parameter family of different theories.

In this paper we shall take a different perspective which
preserves the commutativity of the space-time. In our approach a
fundamental role is played by the unitarily inequivalent
representations (UIRs) of the canonical commutation relations
(CCRs) emerging in the quantization of a field. There are already
well known results obtained in this context: \\
 $i)$ it has been shown the relevance of the UIRs of the CCRs
 in the quantization of a field in a curved background
 \cite{SOD,WALD,FUL}. \\
 $ii)$ on a very general ground, the deformation
 parameter $q$ of the deformed CCRs always labels the UIRs
 for a (relativistic as well as non-relativistic) quantum system with
 infinite number of degrees of freedom \cite{VON}. \\
 $iii)$ thermal properties at equilibrium of a quantum
 field are conveniently described by Thermo Field Dynamics (TFD).
 There quantum algebras provide the natural setting for the
 implementation of the formalism (doubling of the degrees of
 freedom, tilde conjugation, etc.) \cite{Siena}. \\
 $iv)$ it is well known the intimate relationship between space--times
 with an event horizon and thermal properties \cite{ISR,Sanchez}. In
 particular in Ref. \cite{Sanchez} it has been shown that global thermal
 equilibrium over the whole space--time implies the presence of
 horizons in this space--time.

The paper is organized as follows. In Section II we shall review
some mathematical details of the $q$-deformation. We start with a
bosonic system with one degree of freedom, then we move to the
case of the infinite degrees of freedom of a complex scalar field
in Minkowski space--time. Section III is devoted to the
construction of new canonical operators and their Hilbert-Fock
space, naturally arising after deformation. In Section IV we show
how these structures lead to the quantization of complex scalar
fields in static and stationary curved space--times. In Section V
applications of the obtained results are carried out for the
Schwarzschild and Rindler space-times. Conclusions are drawn in
Section VI. Some details of calculations are reported in
Appendices A and B.

\section{Deformation of the CCRs}
\setcounter{equation}{0} Let us consider a complex massive scalar
 quantum field $\phi(x)$ in $n$-dimensional Minkowski space--time
 ($M$-frame), with Lagrangian density
 \begin{equation}\label{lagrangian}
   {\cal L}(\phi^*, \phi)=\partial_\mu \phi^*\partial^\mu
 \phi-m^2\phi^*\phi\,.
 \end{equation}
For sake of simplicity we shall keep the volume $V$ finite
throughout the computation.

As usual, the quantized field $\phi(x)$ can be decomposed in
Minkowski modes $\{U_k(x)\}$, orthonormal with respect to the
Klein-Gordon inner product
 \begin{equation}\label{2}
 \phi(x)=\sum_{k} \,
 [a_{k}U_k(x)\,+\,\bar{a}_k^{\dagger}U_k^{*}(x)] \;,
 \end{equation}
where $k=(k_1, \vec{k})\in {\mathbf{Z}}^{n-1}$ is treated as an
integer label due to the finite volume. The Hamiltonian operator
is then given by
 \begin{equation}
 H_M=\sum_k\,\omega_k\,
 (a_k^{\dagger}a_k\,+\,\bar{a}_k^{\dagger}\bar{a}_k) \;,
 \end{equation}
where $\omega_k= \sqrt{k_1^2+|\vec{k}|^2+m^2}$, and $a_k,
a^\dagger_k$  ($\bar{a}_k,\bar{a}_k^{\dagger}$) are the
annihilation and creation operators, respectively, for particles
(antiparticles). They act on the Fock space $\cal H$, the
Minkowski vacuum being defined by
 \begin{equation}\label{4}
 a_k|0_M>\,=\,\bar{a}_k|0_M>\,=0, \quad \forall \,k\,{.}
 \end{equation}
The operators entering the standard expansion (\ref{2}) satisfy
the usual CCRs
 \begin{equation}
 [a_k, a_{k'}^\dag]=\delta_{kk'}\,, [\bar{a}_k,
 \bar{a}_{k'}^\dag]=\delta_{kk'} \,,
 \end{equation}
 \begin{equation}
 [a_{k} , a_{k'}] = [\bar{a}_k, \bar{a}_{k'}] = 0 \,, \quad \forall
 \,k\,,\,k'\,,
 \end{equation}
which are the conditions for the quantization of the field
$\phi(x)$, in the $M$-frame. We now want to introduce the
deformation of this algebra.

 \subsection{Deformation for a single mode}
Let us consider one set of creation and annihilation operators for
a single mode, for instance $a, \, a^\dagger$. Of course, the same
analysis applies to $\bar{a}, \, \bar{a}^\dagger$.

On a very general ground  $\{ a, \, a^\dagger, \, N, \, c \}$ form
the Weyl-Heisenberg algebra if the following commutation relations
are satisfied
 \begin{equation}\label{a}
 [a,a^{\dagger}] = 2c\,, \quad [N,a] = -a\,, \quad  [N,a^{\dagger}]
 = a^{\dagger}\,, \quad [c, \cdot] = 0 \;.
 \end{equation}
For the moment we include a central term $c$. In the fundamental
representation, defined by $c \equiv \frac{1}{2}$ and the Casimir
$C = 2c N - a^\dagger a \equiv 0$, this algebra reduces to the
standard algebra of quantization . Only after specifying the
representation $N$ is no longer an independent generator, and, in
the fundamental one, is given by $N = a^\dagger a$.

The set $\{ a, \, a^\dagger, \, N, \, c \}$ generates the Hopf
algebra $h(1)$ which, besides the ordinary multiplication of
(\ref{a}), is equipped with three more operations: the coproduct,
the counit and the antipode \cite{tjin}. For our purposes the
crucial ingredient of this Hopf algebra is the coproduct $\Delta$
 \begin{equation}\label{co}
 \Delta a = a \otimes I + I \otimes a \;,
 \end{equation}
and similarly for $a^{\dagger}, \, N$ and $c$. $\Delta$ is defined
to be a homomorphism $\Delta(A\cdot B)=\Delta(A)\cdot\Delta(B)$
for any $A$ and $B$. Operators with coproduct given by (\ref{co})
are called {\it primitive}.
The physical meaning of the coproduct is that it provides the
prescription for operating on two modes. For instance, the
familiar operation of addition of the angular momentum $J^i \in
su(2)$, $i = 1,2,3$, of two particles is a coproduct:
 \[
 \Delta J^i = J^i \otimes I + I \otimes J^i = J_1^i +J_2^i \,.
 \]
In other words, the natural assumption of the additivity of basic
observables, such as the energy and the angular momentum,
necessarily implies to consider the coproduct operation, namely
the Hopf algebra structure.

The deformation of the universal enveloping algebra of $h (1)$,
denoted by $h_q (1)$ \cite{TARL}, is given by
 \begin{equation}\label{aq}
  [a_q,a_q^{\dagger}] = [2c]_q\,, \quad
  [N,a_q] = -a_q\,, \quad [N,a_q^{\dagger}] = a_q^{\dagger}\,, \quad [c,
  \cdot] = 0 \;,
 \end{equation}
where $N$ and $c$ are the same as in $h(1)$ and primitive, but the
commutator of $a_q$ and $a_q^{\dagger}$ is different from the
commutator of $a$ and $a^\dagger$ for generic $c$ (recall that
$[x]_q \equiv \displaystyle{\frac{q^x - q^{-x}}{q - q^{-1}}}$).
Hence for the coproduct of $a_q$ and $a_q^\dagger$ we obtain
 \begin{eqnarray}\label{coq}
 \Delta a_q &=& a_q \otimes q^{c} + q^{-c} \otimes a_q\,, \\
 \Delta a_q^{\dagger} &=& a_q^{\dagger} \otimes q^{c}
                      + q^{-c} \otimes a_q^{\dagger} \,. \nonumber
 \end{eqnarray}
where we use the property $[\Delta a_q, \Delta a_q^\dagger
]=\Delta([a_q, a_q^\dagger])$. The parameter $q$ is called the
{\it deformation parameter}, and in what follows we shall consider
the case of real $q$. We shall later show that many structures
arising in the deformed context acquire a precise physical meaning
when we allow for the deformation parameter to be specified in
terms of suitable physical quantities.

The algebras $h_q (1)$ and $h (1)$  become isomorphic in the
fundamental representation $c=\frac{1}{2}$ (note that $[1]_q =
1$). We shall work in that representation from now on, and that
will have two effects: first the algebra (\ref{a}) reduces to the
standard CCRs of quantization; second the only difference between
$h(1)$ and $h_q (1)$ is now in the coproduct. Thus we shall omit
the label $q$ for the operators, and it will only appear in
$\Delta_q$. Of course the two algebras act on the same
representation space, the Hilbert-Fock space $\cal H$, and both
$\Delta$ and $\Delta_q$ map operators acting on $\cal H$ to
operators acting on $\cal H \otimes \cal H$.

 \subsection{Deformation for a bosonic quantum field}
The next step is to consider the algebraic structure of the entire
field, i.e. to take into account the infinite number of degrees of
freedom labelled by the momentum index $k$. As in the previous
case, we shall focus on one set of operators, for instance $\{a_k,
\, a_k^\dag, \, N_k \}$. The set relative to the antiparticles has
to be treated the same way.

When we deal with the infinite number of degrees of freedom we
have an infinite number of copies of the algebra $h(1)$, one for
each momentum $k$,
 \begin{equation}\label{ak}
 [a_k,a_k^{\dagger}] = 1\,, \quad [N_k,a_k] = -a_k\,, \quad
 [N_k,a_k^{\dagger}] = a_k^{\dagger}\,, \qquad \forall k\,,
 \end{equation}
and different copies $h_k (1)$ and $h_{k'} (1)$, $k \ne k'$, are
related by the standard commutation relations for a quantized
field
 \begin{equation}\label{aa'}
 [a_k,a_{k'}^{\dagger}] = \delta_{k k'}\,,  \quad [a_k,a_{k'}] =
 [a_k^{\dagger},a_{k'}^{\dagger}] = 0\,, \qquad \forall k, k'\,.
 \end{equation}
The deformation of each of the $h_k (1)$s is exactly the same as
for the single mode. We have then just an infinite set of deformed
algebras labelled by $k$.

If one now looks for the relation among the deformed copies of
$h(1)$ with different $k$s, one realizes that the deformation
parameter $q$ can be momentum dependent. If we simply set $ h_k
(1) \rightarrow h_{q(k)} (1) $, a possible generalization of
(\ref{aa'}) is
 \begin{equation}\label{aa'q}
 [a_{q(k)} , a_{q(k')}^\dag] = [\delta_{k k'}]_{q(k)} = \delta_{k
 k'}\,, \qquad \forall k, k'\,,
 \end{equation}
similarly for the other commutators.

More generally, the deformation parameter $q$ could depend on a
momentum $p$ which may or may not coincide with $k$. Furthermore,
$q$ may depend also on other parameters: $q=q($phys$,p)$ (where
``phys'' stands for the proper physical quantity: damping
constant, temperature, etc.). This happens because $q$ labels the
UIRs of the CCRs for quantum fields, as shown in Refs.
\cite{VON,Siena}. We shall keep the shorthand notation $q(p)$
baring the last comments in mind.

 \section{New Canonical Operators and Hilbert Spaces}
 \setcounter{equation}{0}
Let us now define the coproduct of the operators $a_k$ and
$\bar{a}_k$

 \begin{eqnarray}
 \Delta a_k &=& a_k \otimes I + I \otimes a_k \equiv
 a_k^{(+)} + a_k^{(-)}  \;, \nonumber \\
 \Delta {\bar a}_k &=& {\bar a}_k \otimes I + I \otimes {\bar a}_k
 \equiv {\bar a}_k^{(+)} + {\bar a}_k^{(-)} \;. \label{copund}
 \end{eqnarray}
The operators $a_k^{(\sigma)}$ and $\bar{a}_k^{(\sigma)}$,
$\sigma=\pm$, satisfy the CCRs
 \begin{equation}\label{CCRsdouble}
   [a_k^{(\sigma)}, a_{k'}^{(\sigma ')}]=0\,, \quad
   [a_k^{(\sigma)}, a_{k'}^{(\sigma ')\,
   \dagger}]=\delta_{\sigma\sigma'}\delta_{kk'}\,, \quad \,
 \sigma\,,\sigma' = \pm\,,\,\,
   \forall \,k\,,\,k'\,,
 \end{equation}
and similarly for $\bar{a}_k^{(\sigma)}$. They act on the complete
Hilbert space, i.e. ${\cal H}\otimes {\cal H}$, where the ground
state (vacuum) is defined as $\vert 0_M>\otimes \vert 0_M>$. For
brevity we shall indicate it with $\vert 0_M>$. The
$q$-deformation of the coproduct (\ref{copund}) is
 \begin{eqnarray}\label{coqp}
 \Delta_{q} a_k &=& a_k^{(+)} q^{\frac{1}{2}}
                     + a_k^{(-)} q^{-\frac{1}{2}}\,, \nonumber \\
 \Delta_{q} {\bar a}_k &=& {\bar a}_k^{(+)} q^{\frac{1}{2}}
                     + {\bar a}_k^{(-)} q^{-\frac{1}{2}}\,,
 \end{eqnarray}
where $q=q(p)$. Let us stress again that the deformed and
undeformed operators both close under the same algebraic
relations, and the difference shows up only in the coproduct
$\Delta_q$.

We make here some comments. We observe that in the undeformed case
the coproduct (\ref{copund}) maps $a_k \rightarrow a_k^{(+)} +
a_k^{(-)}$ which is a purely formal doubling due to the complete
co-commutativity. The two sets $\{ a_k^{(+)} \}$ and $\{ a_k^{(-)}
\}$ are just identical copies of quantized modes for two fields
$\phi^{(+)}$ and $\phi^{(-)}$ having exactly the same properties.
Something new happens, instead, when we deform. In (\ref{coqp})
the co-commutativity is lost and this means that we obtain two
sectors, $(+)$ and $(-)$, which now are essentially distinct. We
should now ask whether it is possible to build new canonical
operators by using the operators in (\ref{coqp}) and what is their
physical meaning.

In order to preserve the canonical algebra we use a complete
orthonormal set of functions $\{F (k,p)\}$ to introduce the
``smeared" operators
 \begin{equation}\label{standard}
  d_p^{(\sigma)} \equiv \sum_k F (k,p)\, a_k^{(\sigma)}\,, \quad
 {\bar d}_p^{(\sigma)} \equiv \sum_k F (k,p)\, {\bar
 a}_k^{(\sigma)} \;,
 \end{equation}
where $\sigma = \pm$, $p \in {\mathbf{Z}}^{n-1}$, as for $k$, and
$p = (\Omega , \vec{p})$. Strictly speaking one should write the
operators in (\ref{standard}) as $d_{F(p)}^{(\sigma)} \equiv
\sum_k F(k, p)a_k^{(\sigma)} $ and $ {\bar d}_{F(p)}^{(\sigma)}
\equiv \sum_k F (k,p) {\bar a}_k^{(\sigma)} $, as in Ref.
\cite{peppinoF}. However we prefer the simplified notation of
(\ref{standard}). By means of Eq.(\ref{coqp}), the $q$-deformed
coproduct of $d$ and $\bar{d}$ is given by
 \begin{eqnarray}
 \sum_k F (k,p) \Delta_{q(p)} a_k &\equiv& D_q (p) \equiv
   D^{(+)}_q (p) + D^{(-)}_q (p) \;, \nonumber \\
 \sum_k F (k,p) \Delta_{q(p)} {\bar a}_k &\equiv& {\bar D}_q (p)
 \equiv
   {\bar D}^{(+)}_q (p) + {\bar D}^{(-)}_q (p) \;,
 \end{eqnarray}
respectively, where the following short-hand notation has been
used
  \begin{equation}\label{D(p)}
  D^{(\sigma)}_q (p) \equiv q^{\sigma
  \frac{1}{2}}(p) \, d_p^{(\sigma)} \,, \quad
  {\bar D}^{(\sigma)}_q (p) \equiv q^{\sigma \frac{1}{2}}(p)\,{\bar
  d}_p^{(\sigma)}\,.
  \end{equation}
In order to avoid a privileged direction in the phase space we
impose the following constraint on $q(p)$
\begin{equation}\label{cons2}
  q(p) = q(\tilde{p}) \,,
\end{equation}
where we define and $\tilde{p} =\equiv (\Omega , - \vec{p})$
($\Omega > 0$). Of course, Eq. (\ref{cons2}) means that
$q$-deformation must play the same role on particles and
antiparticles. Furthermore we shall make use of the following
standard relation in quantum algebras.
\begin{equation}\label{cons1}
  q(p) = e^{2 \epsilon (p)} \,,
 \end{equation}
where $\epsilon (p)$ is a real function of $p$, as required by our
choice of real $q(p)$.

By using (\ref{D(p)}) we can simply take suitable linear
combinations to obtain
 \begin{eqnarray}
 d_p^{(\sigma)} (\epsilon) &\equiv& \frac{1}{2}
  (D^{(\sigma)}_q (p) + D^{(\sigma)}_{q^{-1}} (p)) +
 \frac{1}{2} ({\bar D}^{(-\sigma) \dagger}_{q^{-1}} (\tilde{p})
              - {\bar D}^{(-\sigma) \dagger}_q  (\tilde{p})) \nonumber \\
  & = & d_p^{(\sigma)} \cosh \epsilon (p)
       + {\bar d}_{\tilde p}^{(-\sigma) \dagger} \sinh \epsilon
       (p)\,,
 \label{b1}
 \end{eqnarray}
and
 \begin{eqnarray}
 {\bar d}_{\tilde p}^{(-\sigma) \dagger} (\epsilon) &\equiv&
 \frac{1}{2}
  (D^{(\sigma)}_q (p) - D^{(\sigma)}_{q^{-1}} (p)) +
 \frac{1}{2} ({\bar D}^{(-\sigma) \dagger}_{q^{-1}} (\tilde{p})
              + {\bar D}^{(-\sigma) \dagger}_q  (\tilde{p})) \nonumber \\
  & = & d_p^{(\sigma)} \sinh \epsilon (p)
       + {\bar d}_{\tilde p}^{(-\sigma) \dagger} \cosh \epsilon
       (p)\,.
 \label{b2}
 \end{eqnarray}
Eqs. (\ref{b1}) and (\ref{b2}) are recognized to be the Bogolubov
transformations. We then succeeded in finding new canonical
operators by using the deformed coproducts (\ref{coqp}).

Let us now look at the Hilbert spaces involved. Again we shall
keep the short-hand notation for the Hilbert spaces (${\cal H}$
stands for $\cal H \otimes \cal H$), as well as for the states
(for instance $|0_M>$ stands for $|0_M> \otimes |0_M>$).

We introduce the infinitesimal generators of the transformations
(\ref{b1}) and (\ref{b2})
 \begin{equation}\label{eq16}
 g(\epsilon)= \sum_p \epsilon(p) g_p =
 \sum_p \sum_{\sigma} \epsilon(p)
 [d_p^{(\sigma)}\bar{d}_{\tilde p}^{(-\sigma)} -
 d_p^{(\sigma)\,\dagger}\bar{d}_{\tilde
 p}^{(-\sigma)\,\dagger}] \,{.}
 \end{equation}
$G(\epsilon) \equiv \exp g(\epsilon)$ is a unitary operator at
finite volume:
$G^{-1}(\epsilon)=G(-\epsilon)=G^{\dagger}(\epsilon)$ and the
Bogolubov transformations (\ref{b1}) and (\ref{b2}) can be now
written as
 \begin{equation}\label{17}
 d_p^{(\sigma)}\to
 d^{(\sigma)}_p(\epsilon)=G(\epsilon)d_p^{(\sigma)}G^{-1}
 (\epsilon)\,{,}
 \end{equation}
 \begin{equation}\label{18}
 \bar{d}_{\tilde p}^{(-\sigma)\dagger}\to \bar{d}_{\tilde
 p}^{(-\sigma)\dagger}(\epsilon)= G(\epsilon)\bar{d}_{\tilde
 p}^{(-\sigma)\dagger}G^{-1} (\epsilon)\,{.}
 \end{equation}
The Hilbert--Fock space $\cal  H$ of the basis vectors associated
to the Minkowski space  is build by repeated action of
$(d_p^{(\sigma)\,\dagger}, \bar{d}_{\tilde{p}}^{(-\sigma)\,\dag})$
on the vacuum state $|0_M>$. The Bogolubov transformations with
parameter $\epsilon$, Eqs. (\ref{17}) and (\ref{18}), relate
vectors of $\cal H$ to vectors of another Hilbert space ${\cal
H}_{\epsilon}$ labelled by $\epsilon$. The relation between these
spaces is established by the generator $G(\epsilon)$: ${\cal H}
\to {\cal H}_{\epsilon}$. In particular, for the vacuum state
$|0_M>$ one has
 \begin{equation}\label{19}
 |0(\epsilon)>\,=\,G(\epsilon)\,|0_M>\,{,}
 \end{equation}
where $|0(\epsilon)>$ is the vacuum state of the Hilbert space
${\cal H}_{\epsilon}$ annihilated by the new operators
($d_p^{(\sigma)} (\epsilon)$ , $\bar{d}_{\tilde
p}^{(-\sigma)}(\epsilon)$). Note that even though $\epsilon =
\epsilon({\rm phys} , p)$, the vacuum $|0(\epsilon)>$ does not
depend on $p$, but only on the physical parameter, as can be
easily seen from Eq. (\ref{eq16}).

The group underlying this construction is $SU(1,1)$, as can be
easily seen, for instance, by noticing that $g(\epsilon)$ is one
of its generators. Thus, by inverting Eq. (\ref{19}), and using
the Gaussian decomposition for the group $SU(1,1)$, the Minkowski
vacuum can be expressed as a $SU(1,1)$ generalized coherent state
\cite{PER} of Cooper-like pairs
 \begin{equation}\label{26}
 |0_M>=\frac{1}{Z}\,\exp\left[{\sum_{\sigma} \sum_p \;\tanh\epsilon (p)
 d_p^{(\sigma)\dagger}(\epsilon) \bar{d}_{\tilde
 p}^{(-\sigma)\dagger}}(\epsilon)\right]\, |0(\epsilon)>\,{,}
 \end{equation}
where $Z= \prod_p\;\cosh^2\epsilon(p)$.

Moreover, $<0(\epsilon)|0(\epsilon)>=1, \forall \epsilon$, and in
the infinite-volume limit, we have
  \begin{eqnarray}
  <0(\epsilon)|0_M>\to 0 & &\quad {\rm as} \quad  V\to\infty, \quad \forall
 \epsilon
 \label{eq27} \\
  <0(\epsilon)|0(\epsilon^{\prime})>\to 0 & &\quad {\rm as} \quad
   V\to\infty, \quad \forall
  \epsilon, \epsilon^{\prime}, \epsilon\ne \epsilon^{\prime}\,{,}
  \label{eq28}
  \end{eqnarray}
i.e. the Hilbert spaces ${\cal  H}$ and ${\cal H}_\epsilon$ become
unitarily inequivalent in the infinite volume limit ($V$-limit).
In this limit $\epsilon$ labels the set $\{H_\epsilon, \forall
\epsilon\}$ of the infinitely many UIRs of the CCRs.

 \section{Thermal Properties and Space-Time with Event Horizons}
 \setcounter{equation}{0}
The formalism constructed in the previous Sections presents some
common features with the formalism of quantum dissipation in the
approach of \cite{FREE,VON}. In those papers the UIRs  of the CCRs
for a field of quantized damped harmonic oscillators are labelled
by $\Gamma t$, with $\Gamma$ the damping constant and $t$ the time
variable. Thus the Hilbert spaces ${\cal H}$ and ${\cal
H}_\epsilon \equiv {\cal H}_{\Gamma t}$, for any $t$, have the
Minkowski space--time as support. This is made possible by
identifying in (\ref{coqp}) the momenta $k$ and $p$, but, of
course, this is not the only possible choice. In fact, one could
explore the more general case in which the momenta are distinct,
as we shall do in this paper.

The physical meaning of having two distinct momenta $k$ and $p$
for states in the Hilbert spaces ${\cal H}$ and ${\cal
H}_\epsilon$, respectively, is the occurrence of two {\it
different} reference frames: the $M$-frame (Minkowski) and the new
frame, which we shall call $M_\epsilon$-frame. To connect them one
introduces the generator of the boosts. For instance, let us
consider the $(1,0)$--component of the generator of the Lorentz
transformations defined as \cite{ZUB}
 \begin{equation}\label{boost}
 M_{10}= - i \sum_k \left[a^{\dagger}_{k_1\vec{k}} \sqrt{\omega_k}
 \left(\frac{\partial}{\partial k_1} \sqrt{\omega_k}\,
 a_{k_1\vec{k}}\right) + (a \rightarrow \bar{a})\right]\,{.}
 \end{equation}
The structure to be considered is the coproduct of $M_{10}$,
 \begin{eqnarray}\label{boostdef}
  {\cal M}_{10} &=& \Delta M_{10}= M_{10}\otimes I+I\otimes M_{10}
  \\
  &=&  - i \sum_k \left[a^{(+)\,\dagger}_{k_1\vec{k}} \sqrt{\omega_k}
  \left( \frac{\partial}{\partial k_1} \sqrt{\omega_k}\,
  a_{k_1\vec{k}}^{(+)}\right) + (a_k^{(+)} \rightarrow
  \bar{a}_k^{(+)})\right. \nonumber \\
 & & +\left. a^{(-)\,\dagger}_{k_1\vec{k}} \sqrt{\omega_k}
  \left(\frac{\partial}{\partial k_1} \sqrt{\omega_k}\,
  a_{k_1\vec{k}}^{(-)}\right) + (a_k^{(-)} \rightarrow \bar{a}_k^{(-)})
 \right]\,{.} \nonumber
 \end{eqnarray}
In the Appendix A we shall prove that ${\cal M}_{10}$ is the only
admissible expression within our formalism. This happens because
the generators of the Poincar\'e algebra are primitive under the
{\it same} coproduct $\Delta$ of the undeformed algebra $h(1)$ for
the quantized scalar field $\phi (x)$. Note that in the deformed
case things are not so simple. The $q$-deformation of $h(1)$ is
non-linear, and this imply that we shall have different deformed
coproducts for $h_q(1)$ and the deformed Poincar\'e $ISO_q(3,1)$.

To explore the physics in the $M_\epsilon$--frame, one has to
construct a diagonal operator which plays the role of the
Hamiltonian. We start by inverting the expressions
(\ref{standard}) and then we substitute them in (\ref{boostdef}).
With our choice for the boost the only relevant components of the
momenta $k = (k_1, \vec{k})$ and $p = (\Omega, \vec{p})$ are $k_1$
and $\Omega$. With no loss of generality, we can equal the other
components, $\vec{k}\equiv \vec{p}$. In the Appendix B we shall
prove that, in order to diagonalize ${\cal M}_{10}$ in terms of
$d_p^{(\sigma)}$ and $\bar{d}_{\tilde p}^{(\sigma)}$, we have to
demand the complete set of functions $\{F(k_1, \Omega)\}$ to
satisfy the following equation in the two sectors $\sigma=\pm$
 \begin{equation}\label{feqdiff}
   \frac{\partial}{\partial k_1}\left[ \sqrt{\omega_k} \,
   F(k_1, \Omega)\right] =
   i \sigma\,\Omega \, \frac{1}{\sqrt{\omega_k}} \,
   F(k_1, \Omega)\,.
 \end{equation}
Therefore one obtains
 \begin{equation}\label{M10diag}
  {\cal M}_{10}=  \sum_{\sigma} \sum_p \,\sigma\, \Omega
 \,[ d_p^{(\sigma)
 \dagger} d_p^{(\sigma)} + \bar{d}_{\tilde{p}}^{(\sigma)}
 \bar{d}_{\tilde{p}}^{(\sigma)\dagger}]\,{.}
 \end{equation}
This does not happen for the $a$ and $\bar{a}$ operators we
started with in the expansion (\ref{2}). As shown in Appendix B,
the differential equation (\ref{feqdiff}) is solved in the two
sectors $\sigma=\pm$ by
  \begin{equation}\label{fexpr}
  F(k_1, \Omega)=F_\sigma (k_1, \Omega)=\frac{1}{\sqrt{2\pi\omega_k}}\left(
  \frac{\omega_k+k_1}{\omega_k-k_1}\right)^{i\sigma\Omega/2} \,{,}
  \end{equation}
where one has to take the solution with $\sigma =+$ for the sector
where $d^{(+)}_{\Omega \vec{k}}$ (or $a^{(+)}_{k_1\vec{k}}$) acts
and $\sigma =-$ for the sector where $d^{(-)}_{\Omega \vec{k}}$
(or $a^{(-)}_{k_1\vec{k}}$) acts. We then succeed in diagonalizing
${\cal M}_{10}$ in terms of operators which act on ${\cal H}$, in
the $M$-frame. Now it is a simple matter to express ${\cal
M}_{10}$ in terms of the new operators $d^{(\sigma)}_p(\epsilon)$
and $\bar{d}^{(\sigma)}_{\tilde p} (\epsilon)$ given in Eqs.
(\ref{17}) and (\ref{18}), respectively:
 \begin{eqnarray}
  H_{\epsilon}&=&G(\epsilon){\cal M}_{10}G^{-1}(\epsilon) \nonumber \\
  & = & \sum_{\sigma}\sum_p \sigma \Omega \,[ d_p^{(\sigma)
 \dagger}(\epsilon) d_p^{(\sigma)}(\epsilon) +
 \bar{d}_{\tilde{p}}^{(\sigma)}(\epsilon)
 \bar{d}_{\tilde{p}}^{(\sigma)\dagger}(\epsilon)] \nonumber \\
 &=& H^{(+)}(\epsilon) - H^{(-)}(\epsilon) \,{.}
 \label{hamrin}
 \end{eqnarray}
A comment is in order: In (\ref{feqdiff}) (see also (\ref{fexpr}))
we introduce two sets of functions, one for each sector labelled
by $\sigma=\pm$, thus the Cooper-like pairs in the vacuum
(\ref{26}) have zero energy.
It is important to notice that $ H_{\epsilon}$ acts on states
defined in the {\it new} Hilbert space ${\cal H}_\epsilon$, and
has as space--time the new frame $M_\epsilon$. Moreover,
$H_\epsilon$ is the $SU(1,1)$ Casimir operator. Note also that
this Hamiltonian is correctly expressed in terms of number
operators for the modes $d_p^{(\sigma)}(\epsilon)$ and
$\bar{d}_{\tilde p}^{(\sigma)}(\epsilon)$. From Eqs. (\ref{17})
and (\ref{18}) one immediately realizes that the number of modes
of the type $d_p^{(\sigma)}(\epsilon)$ in $|0_M>$ is given by
 \begin{equation}\label{25}
 {\cal N}^{(\sigma)}_{d(\epsilon)} \equiv <0_M|d_p^{(\sigma)
 \dagger}(\epsilon) d_p^{(\sigma)}(\epsilon)|0_M>=
 \sinh^2\epsilon(p) \;, \quad \sigma = \pm \;,
 \end{equation}
and similarly for the modes of type $\bar{d}_{\tilde p}^{(\sigma)}
(\epsilon)$.

Due to the condensate structure of the vacuum (\ref{26}), we can
consider the thermal properties of the quantum system described by
operators (\ref{17}) and (\ref{18}) by introducing the entropy
operator.

At this end we recast the vacuum state $|0_M>$, given in Eq.
(\ref{26}), in the form \cite{umezawa1}
\begin{eqnarray}\label{vacua+entropy}
  |0_M> &=& e^{-S^{(+)}(\epsilon)/2}\,
        e^{\sum_{p, \sigma}d^{(\sigma)\, \dagger}_p(\epsilon)\,
        {\bar d}^{(-\sigma)\, \dagger}_{\tilde
        p}(\epsilon)}|0(\epsilon)>
        \\
        &=& e^{-S^{(-)}(\epsilon)/2}\,
        e^{\sum_{p, \sigma}d^{(\sigma)\, \dagger}_p(\epsilon)\,
        {\bar d}^{(-\sigma)\, \dagger}_{\tilde
        p}(\epsilon)}|0(\epsilon)>\,,
        \nonumber
\end{eqnarray}
where $S^{(+)}(\epsilon)$ and $S^{(-)}(\epsilon)$ are given by
 \begin{eqnarray}\label{S+}
 S^{(+)}(\epsilon)&=& {\cal S}^{(+)}(\epsilon) + \bar{\cal
 S}^{(+)}(\epsilon) \\
 &=& - \sum_p [d_p^{(+) \dagger}(\epsilon) d_p^{(+)}(\epsilon)
  \ln\sinh^2\epsilon (p)
 - d_p^{(+)}(\epsilon)d_p^{(+)\dagger}(\epsilon) \ln\cosh^2\epsilon
 (p) \nonumber \\
 & & + (d \to \bar{d})] \,, \nonumber
 \end{eqnarray}
 \begin{eqnarray}\label{S-}
 S^{(-)}(\epsilon)&=& {\cal S}^{(-)}(\epsilon) + \bar{\cal
 S}^{(-)}(\epsilon) \\
 &=& - \sum_p [d_p^{(-) \dagger}(\epsilon) d_p^{(-)}(\epsilon)
  \ln\sinh^2\epsilon (p)
 - d_p^{(-)}(\epsilon)d_p^{(-)\dagger}(\epsilon) \ln\cosh^2\epsilon
 (p) \nonumber \\
 & & + (d \to \bar{d})] \,. \nonumber
 \end{eqnarray}
To prove (\ref{vacua+entropy}) one must use the following very
useful relations \cite{umezawa1}
\begin{eqnarray}\label{relaz1}
 e^{-S^{(+)}(\epsilon)/2}\, d^{(+)\dagger}_p(\epsilon)
 e^{S^{(+)}(\epsilon)/2}
 &=&\tanh \epsilon(p) \,d^{(+)\dagger}_p(\epsilon) \,,\\
 e^{-S^{(+)}(\epsilon)/2}\, \bar{d}^{(+)\dagger}_p(\epsilon)
 e^{S^{(+)}(\epsilon)/2}
 &=&\tanh \epsilon(p) \,\bar{d}^{(+)\dagger}_p(\epsilon) \,, \nonumber
\end{eqnarray}
similarly for $d^{(-) \dagger}_{\tilde p}$ and ${\bar d}^{(-)
\dagger }_{\tilde p}$ by using $S^{(-)}(\epsilon)$, and
\begin{equation}\label{relaz2}
 e^{-S^{(+)}(\epsilon)/2}\,|0(\epsilon)>=
 e^{-S^{(-)}(\epsilon)/2}\,|0(\epsilon)> =
 \prod_p\,\cosh^{-2} \epsilon(p)|0(\epsilon)>\,.
\end{equation}
The operator $S^{(+)}(\epsilon) = {\cal S}^{(+)}(\epsilon) +
\bar{\cal S}^{(+)}(\epsilon)$ is the sum of the entropy operators
for the (free) boson gas of particles and antiparticles in the
sector $(+)$ of the $M_\epsilon$-frame, similarly for
$S^{(-)}(\epsilon)$. The total entropy operator is given by
\begin{equation}\label{Stotale}
S_\epsilon = S^{(+)}(\epsilon) - S^{(-)}(\epsilon) \,,
\end{equation}
and, as the Hamiltonian (\ref{hamrin}), it is the {\it difference}
of the two operators. The Bogolubov transformations leave
$S_\epsilon$ invariant, $[S_\epsilon, g(\epsilon)]=0$, hence the
relation
\[
S_\epsilon |0_M> = 0
\]
holds. This means that one can arbitrarily choose one of the two
sectors of the $M_\epsilon$-frame, $\sigma = \pm $, to ``measure"
the correspondent entropy $S^{(\pm)}(\epsilon)$ relative to the
ground state of the $M$-frame $|0_M>$. Let us work in the sector
$\sigma = +$.

One can easily convince himself that $S^{(+)}(\epsilon)$ is the
well known von Neumann entropy
 \begin{equation}\label{vonNeuE}
   S = - {\cal N} \ln {\cal N} \,,
 \end{equation}
where ${\cal N}$ is the number of microscopic states. By counting
the number of occupation states in the condensate $|0_M>$ with the
number operator for particles $N^{(+)}_{d(\epsilon)} \equiv
d_p^{(+) \,\dagger}(\epsilon) d_p^{(+)}(\epsilon)$, we must
subtract those occupation states counted by the operator
$\bar{d}_{\tilde p}^{(+)}(\epsilon) \bar{d}_{\tilde p}^{(+)\,
\dagger}(\epsilon) = 1 + N^{(+)}_{\bar{d}(\epsilon)}$, where
$N^{(+)}_{\bar{d}(\epsilon)}$ is the number operator for the
antiparticles. It follows, then, our definition (\ref{S+}) of the
entropy operator
 \begin{eqnarray}\label{23}
   S &=& - \sum_p [N^{(+)}_{d(\epsilon)}
 \ln {\cal N}^{(+)}_{d(\epsilon)} - (1 + N^{(+)}_{\bar{d}(\epsilon)}) \ln (1 +
 {\cal N}^{(+)}_{\bar{d}(\epsilon)}) \\
 & & + (d \to \bar{d})] \nonumber \\
 &\equiv& S^{(+)}(\epsilon) \,. \nonumber
 \end{eqnarray}
Let us now introduce the free-energy defined as
\cite{umezawa1,umezawa,FREE}
 \begin{equation}\label{22}
  {\cal F}^{(+)}(\epsilon)\equiv <0_M|H^{(+)} (\epsilon)
  -\frac{1}{\beta} \, S^{(+)}(\epsilon)|0_M> \,{.}
 \end{equation}
$\beta$ is a strictly positive function to be determined. Looking
for the values of $\epsilon (p)$ making ${\cal F}^{(+)}(\epsilon)$
stationary, one obtains
 \begin{equation}\label{24}
 \beta\Omega = - \ln\tanh^2\epsilon (p) \Leftrightarrow
 \sinh^2\epsilon (p)= \frac{1}{e^{\beta\Omega}-1}\,{,}
 \end{equation}
from which, by using the definition in Eq. (\ref{25}), follows
that
 \begin{equation}\label{Ndist}
 {\cal N}^{(+)}_{d(\epsilon)} = \frac{1}{e^{\beta\Omega}-1}\,,
 \end{equation}
and similarly for ${\cal N}^{(+)}_{\bar{d}(\epsilon)}$.

We thus see that the number of particles ${\cal
N}^{(+)}_{d(\epsilon)}$ in the $M_\epsilon$-frame computed in the
Minkowski vacuum $|0_M>$ gives a Bose-Einstein distribution
provided $\beta$ is identified with the inverse of the
temperature: $\beta = T^{-1}$. We stress here that Eq.
(\ref{Ndist}) is a direct consequence of the interplay between the
doubling ($d^{(+)}(\epsilon)$ and $d^{(-)}(\epsilon)$, and
similarly for the antiparticles) built in the coproduct, and the
deformation of the latter. The induced partition of the
$M_\epsilon$ space--time into two sectors, $\sigma = +$ and
$\sigma = -$, indicates then the emergence of an event horizon,
namely the emergence of a gravitational field. All that is encoded
in the condensate structure of the vacuum realized in Eq.
(\ref{Ndist}).

In order to have a better understanding of how the event horizon
naturally arises in our approach let us recall that in the sector
$\sigma = +$ one has to use only $H^{(+)}(\epsilon)$, while in the
sector $\sigma = -$ one has to use only $H^{(-)}(\epsilon)$.
Therefore there is a boundary region, say $\Sigma$, where for
consistency we must have
\[ H^{(+)}|_\Sigma = H^{(-)}|_\Sigma
\]
so that the event horizon is characterized as the region of
$M_\epsilon$ where one must use the full Hamiltonian $H_\epsilon$
given in Eq. (\ref{hamrin}), and on the vacuum $|0(\epsilon)>$
this will give zero. Of course the same argument applies to the
entropy.

Summarizing, from Eq. (\ref{24}) one sees that for vanishing $T$
the deformation parameter $\epsilon$ vanishes too. In that limit
thermal properties as well as event horizons are lost, and
$M_\epsilon$-frame $\to$ $M$-frame. Moreover, $i)$ $\beta$ is
related to the event horizons, and being $\beta$ constant in time
the $M_\epsilon$ space--time is static and stationary; $ii)$ the
gravitational field itself vanishes as $\epsilon \to 0$. The
 vanishing of the gravitational field occurs either if the
 $M$-frame is far from the gravitational source where space-time is
 flat, or if there exists a reference frame locally flat, i.e. the
 $M$-frame is a free--falling reference frame. This clearly is a realization
 of the equivalence principle, which manifests itself when
 "$\epsilon$-effects" are shielded.

It is now clear the physical meaning of our results: we succeeded
in quantizing a scalar field in a curved background. The CCRs in
the $M_\epsilon$--frame have not to be imposed by hands, as
usually done in QFT in curved space--time, but they are naturally
recovered via $q$-deformation: The physical parameter
characterizing the background geometry is related to the
$q$-deformation parameter.


 \section{Applications to Schwarzschild and Rindler space--times}
 \setcounter{equation}{0}

The aim of this Section is to study the consequences of the
previous results in two specific cases: The Schwarzschild and
Rindler geometries in $(3+1)$-dimensions.

 \subsection{\bf The Schwarzschild space--time}
As we have seen, the parameter $\beta$ is related to the event
horizons, which forbid an observer to acquire knowledge to what
happens beyond it. In General Relativity the event horizon is
determined by the vanishing of the $00$-component of the metric
tensor, $g_{00}(r)=0$. In the Schwarzschild space-time, the event
horizon is given by the Schwarzschild radius $r_S=2GM$, being $M$
the mass of the gravitational source and $G$ the Newtonian
constant. According to the above discussion, we can impose the
identification $\beta\sim r_S$, or $T\sim (GM)^{-1}$, which is the
Bekenstein-Hawking temperature defined in (\ref{1a}). The
$q$-deformation parameter is then related to the Schwarzschild
radius, which characterizes the geometrical structure underlying
the $M_\epsilon$-frame.

Let us now analyze the entropy operator. At the origin of the
entropy (\ref{S+}) (see also (\ref{Stotale}) and (\ref{23})) there
are the vacuum fluctuations of quantum states, which have a
thermal character for different observers related to the Minkowski
observer through a diffeomorphism. In the following calculations
we shall perform the continuum limit, $\displaystyle{\sum_p\to
\frac{V}{(2\pi)^3}\int dp=\frac{V}{(2\pi)^3}\int_0^\infty
d\Omega\int d^2 k}$, and we shall use $\displaystyle{d_p^{(+)\,
\dagger}(\epsilon)|0(\epsilon)>=b^{(+)}(\vec{p})| p, \epsilon>}$,
where $b^{(+)}(\vec{p})$ is a wave--packet ($\displaystyle{\int
|b^{(+)} (\vec{p})|^2 d^2 p <\infty}$), similarly for
$\bar{d}_{\tilde p}^{(+)\, \dagger}(\epsilon)$. By computing the
expectation value on the vacuum $|0_M>$ of Eq. (\ref{S+}), one
obtains the entropy density
 \begin{eqnarray}\label{entropy}
 s^{(+)} (\epsilon) &=& \frac{<0_M| S^{(+)}(\epsilon) |0_M>}{V} \\
& \sim & - \int_0^{\infty}d \Omega \sinh^2 \epsilon(\Omega) \ln
 \sinh^2\epsilon(\Omega)
+ \int_0^{\infty}d \Omega \cosh^2 \epsilon(\Omega)\ln
 \cosh^2\epsilon(\Omega) \,{.} \nonumber
 \end{eqnarray}
The integration in Eq. (\ref{entropy}) gives \cite{ryzhik}
\begin{equation}\label{entrbeta}
s^{(+)} (\epsilon) \sim \beta^{-1} \geq 0\,,
\end{equation}
that means that the entropy is a non-negative function, i.e. never
decreases in agreement with the second law of black holes
thermodynamics.

{}From the Eq. (\ref{entrbeta}) immediately follows that the
entropy $<S^{(+)} (\epsilon)>_M$ of a spherical volume of radius
$r_S$ is
 \begin{equation}\label{entropy1}
 <S^{(+)}(\epsilon)>_M = \frac{4\pi r_S^3}{3} \; s^{(+)}(\epsilon) \sim r_S^2 \sim
 (GM)^2\sim {\cal A} \,{,}
 \end{equation}
namely the entropy is proportional to the horizon area ${\cal A}$
of the black holes \cite{HAW,mitra}.

 \subsection{\bf The Rindler space--time}
The Rindler space-time is associated to a uniformly accelerated
observer (the $M_\epsilon$-frame) \cite{rindler}. Shortly we
recall that the presence of the horizon follows because the
Rindler coordinates cover only two disconnected regions of the
Minkowski space--time, $R_+=\{x^\mu = (x^0, x^1, y_R,
z_R)\,|x^1>|x^0|\}$ and $R_- =\{x^\mu = (x^0, x^1, y_R, z_R
)\,|x^1<-|x^0|\}$, where the components ${\vec x}_R=(y_R, z_R)$
coincide with the Minkowskian ones ${\vec x}_M=(y_M, z_M)$, ${\vec
x}_M\equiv {\vec x}_R$.
From the general relations between the four-velocity ${\dot
x}^\mu$ and the four-acceleration ${\ddot x}^\mu$, ${\dot
x}^\mu{\dot x}_\mu=-1$,  and ${\dot x}^\mu {\ddot x}_\mu=0$, one
derives the relation between the Minkowski and Rindler
coordinates: $x^1= \xi \cosh \eta$ and $x^0= \xi \sinh \eta$,
where $\eta$ is the Rindler time, $\xi$ is the ``spatial"
coordinate. The Rindler coordinates are given by $x_R^\mu = (\eta
\equiv a \tau, \xi \equiv a^{-1}, y_R, z_R)$, where $\tau$ is the
proper time, and $a$ is the constant acceleration of the reference
frame along the $x_R^1$-direction.

The horizon is characterized by $a^{-1}$, and $g_{00}(a)$ vanishes
in the limit $a\to \infty$. Then, from our discussion in Sec. IV,
$\beta\sim a^{-1}$, or equivalently, $T\sim a$, which is the
thermalization theorem Eq. (\ref{1}).

Thus the $q$-deformation parameter turns out to be related to the
acceleration $a$. Such result is formally equivalent to the
Schwarzschild geometry since the surface of gravity of the black
holes in (\ref{1a}) is the gravitational {\it acceleration} at
radius $r$ measured at the infinity.

 {}From (\ref{entrbeta}) one gets $s^{(+)}(\epsilon)
\sim \beta^{-1} \sim a$. The volume is $V=\int d\xi dy_R
dz_R={\cal A} a^{-1}$, being ${\cal A} = \int dy_R dz_R$ the area
of the surface of constant $x_R^1$. Then, the entropy is $S^{(+)}
(\epsilon) \sim {\cal A}$, which confirms (\ref{entropy1}), and is
in agreement with the result of Ref. \cite{laflamme}.

 \section{Conclusions}
 \setcounter{equation}{0}
We have shown how the $q$-deformation of the canonical algebra of
a complex scalar field quantized in the Minkowski space--time
reproduces some of the typical structures of a quantized field in
a space--time with horizon, provided one demands suitable
conditions on the deformation parameter $q(p)$ to be satisfied.

The proper physical quantity on which the deformation parameter
$q$ does depend, is related to the geometrical properties
characterizing the background: For the Schwarzschild geometry this
parameter is the Schwarzschild radius, hence $q=q(r_S)$, whereas
for the Rindler space-time is the acceleration, hence $q=q(a)$.
The parameter $q$(phys)$\to 0$ as the physical parameters vanish,
and the curved space-time reduces to a (locally) flat space-time.
As a consequence, we are led to conclude that quantum deformations
can be induced by gravitational fields. This is the main result of
our work.

Of course, in this paper we have presented only preliminary
results, and surely more work is needed to fully understand the
possible implications this approach could have in handling QFT in
curved space-time, hence in a generic gravitational background.
Nonetheless, it seems to us that this paper is a promising
indication that $q$-deformed algebras provide a natural setting
for treating the quantization of a field in a curved background,
suggesting, among other things, a deep connection between
$q$-groups and quantum gravity, as argued in \cite{MAG}.

\acknowledgments A.I. and G.V. acknowledge the financial support
of the European Science Foundation. A.I. has been partially
supported by the Fellowship ``Theoretical Physics of the
Fundamental Interactions'', University of Rome Tor Vergata, and
acknowledges the kind hospitality of the Dublin Institute for
Advanced Studies. G.L. has been supported by MURST PRIN 99.

 \appendixa

In this Appendix we prove that the generators of the Poincar\'e
algebra (constructed from the energy-momentum tensor $T^{\mu \nu}$
for this theory) are primitive under the same coproduct $\Delta$
of the $h(1)$ algebra of creation and annihilation operators for
the quantized scalar field $\phi (x)$.

The Poincar\'e algebra is given by
 \begin{equation}
 [p^\mu , p^\nu] = 0 \,,\label{a1p}
 \end{equation}
 \begin{equation}
 [{\cal J}^{\mu \nu} , {\cal J}^{\rho \sigma}] = \eta^{\mu \rho} {\cal J}^{\nu \sigma}
  - \eta^{\nu \sigma} {\cal J}^{\mu \rho} + \eta^{\nu \rho} {\cal J}^{\mu \sigma}
  - \eta^{\mu \rho} {\cal J}^{\nu \sigma} \,, \label{a1m}
 \end{equation}
 \begin{equation}
  [{\cal J}^{\mu \nu} , p^{\rho}] = \eta^{\mu \rho} p^\nu - \eta^{\nu
  \rho} p^\mu \,.\label{a1pm}
 \end{equation}
For a given relativistic scalar theory with Lagrangian density
${\cal L}(\phi , \partial \phi)$, the generators of the Poincar\'e
algebra can be expressed as momenta (of order zero and one) of the
conserved energy-momentum tensor
 \begin{equation}\label{a2}
   T^{\mu \nu} = \frac{\partial \cal L}{\partial \partial_\mu
   \phi} \partial^\nu \phi - \eta^{\mu \nu} {\cal L} \,,
 \end{equation}
by defining
 \begin{equation}\label{a3}
   p^\mu \equiv T^{0 \mu} \quad {\rm and} \quad {\cal J}^{\mu \nu} \equiv
   {\cal J}^{0 \mu \nu} \,,
 \end{equation}
where ${\cal J}^{\mu \nu \lambda} = T^{\mu \lambda} x^\nu - T^{\mu
\nu} x^\lambda$.

In our case, by expressing the (complex) scalar field $\phi(x)$ in
terms of quantized modes $a_k , \bar{a}_k$, we have a two-boson
realization of the algebra (\ref{a1p})-(\ref{a1pm}). Namely the
generators of the Poincar\'e algebra are expressed in terms of the
$a_k , \bar{a}_k$. For sake of simplicity let us consider the
particle sector of the theory, and let us denote by $p^\mu(a)$ and
${\cal J}^{\mu \nu}(a)$ the corresponding generators.

A delicate point is how to deal with $p^\mu(a)$ and ${\cal J}^{\mu
\nu}(a)$ when the coproduct is introduced for the underlying
$h(1)$. For instance, one could choose to express $\phi (x)$ in
(\ref{a2}) in terms of $a_k$ and use the fact that $\Delta$ is a
homomorphism ($\Delta (A B) = \Delta (A) \cdot \Delta(B)$), but
this choice is inconsistent and would produce non primitive
$p^\mu(a)$ and ${\cal J}^{\mu \nu}(a)$.

This is easily seen if we consider the operator $N$ of $h(1)$, for
a single mode (the generalization to $k$ modes is trivial) . On
the one hand this operator, regarded as an independent generator
of $h(1)$, is primitive
 \begin{equation}\label{a4}
   \Delta N = N \otimes I + I \otimes N \,.
 \end{equation}
On the other hand, in the fundamental representation identified by
$c=\frac{1}{2}$ and $C=0$, $N = a^\dagger a$ (see also the
comments after (\ref{a})), and its coproduct is given by
 \begin{equation}\label{a5}
   \Delta (a^\dagger a) = a^\dagger a \otimes I + I \otimes a^\dagger a
   + a^\dagger \otimes a + a \otimes a^\dagger \,.
 \end{equation}
The correct way to handle $\Delta N$ is then to use (\ref{a4}) and
{\it only after} that use $N = a^\dagger a$ to write
 \begin{equation}\label{a6}
   \Delta N =  a^{(+)\dagger} a^{(+)} +  a^{(-)\dagger}
 a^{(-)} \,,
 \end{equation}
with the notations introduced in Sec. III.

At this point we use the expression for the Hamiltonian obtained
by Legendre transforming the Lagrangian, and we write
 \begin{equation}\label{a7}
   H = \omega N
 \end{equation}
where $\omega$ is a c-number. This implies that $H \equiv p^0(a)$
is primitive, and, by invoking covariance, $p^\mu(a)$ has to be
primitive as well
 \begin{equation}\label{a8}
 \Delta p^\mu(a) = p^\mu(a) \otimes I + I \otimes p^\mu(a) \,,
 \end{equation}
under the {\it same} coproduct $\Delta$ of $h(1)$. By using
(\ref{a1pm}), and requiring that
 \begin{equation}\label{a9}
 [\Delta {\cal J}^{\mu \nu}(a) , \Delta p^{\rho}(a)] =
   \Delta ([{\cal J}^{\mu \nu} (a), p^{\rho}(a)]) =
   \eta^{\mu \rho} \Delta p^\nu (a)- \eta^{\nu \rho} \Delta p^\mu (a) \,,
 \end{equation}
one immediately obtains that ${\cal J}^{\mu \nu}(a)$ has to be
primitive under the same $\Delta$ of $h(1)$.

We then conclude that the correct expression to consider for the
coproduct of $M^{10} \equiv \sum_k {\cal J}^{10}(a_k, \bar{a}_k)$,
is the one given in (\ref{boostdef}).

 \appendixb

In this Appendix we shall show that the set of functions $\{F(k,
p)\}$ is complete, i.e.
 \begin{eqnarray}\label{appb1}
   \sum_p F^*(k, p)F(k', p)&=&\delta_{kk'} \,, \\
  \sum_k F^*(k, p)F(k, p')&=&\delta_{pp'} \,, \label{appb2}
 \end{eqnarray}
and we shall derive an explicit expression. For this, we consider
only the $k_1$ and $\Omega$ components of the $(n-1)$-vectors
$k=(k_1, \vec{k})$ and $p=(\Omega, \vec{k})$. We shall now perform
the continuum limit. From Eq. (\ref{standard})  it follows
 \begin{equation}\label{b3}
   a_{k_1\vec{k}}^{(\sigma)}=\int dk_1d^{n-2}k F(k_1, \Omega)
   d^{(\sigma)}_{\Omega\vec{k}}\,,
 \end{equation}
and similarly for $\bar{a}_k^{(\sigma)}$. Substituting in
 (\ref{boostdef}) one gets
 \begin{equation}\label{b4}
   {\cal M}_{10}= {\cal M}_{10}^{(+)}+{\cal M}_{10}^{(-)}\,,
 \end{equation}
where
  \begin{eqnarray}\label{b4+}
  {\cal M}_{10}^{(+)}&=& - i \int dk_1d^{n-2}k\int d\Omega 'd\Omega
   d^{(+)\,\dagger}_{\Omega '\vec{k}}\,
  d^{(+)}_{\Omega\vec{k}}\,
  F^*(k_1, \Omega ')\sqrt{\omega_k}
  \left(\frac{\partial}{\partial k_1}
  \sqrt{\omega_k} F(k_1, \Omega)\right) \\
  & &+ (d_p \rightarrow \bar{d}_{\tilde p})\,{,} \nonumber
 \end{eqnarray}
  \begin{eqnarray}\label{b4-}
  {\cal M}_{10}^{(-)}&=& - i \int dk_1d^{n-2}k\int d\Omega 'd\Omega
  d^{(-)\,\dagger}_{\Omega '\vec{k}}\,
  d^{(-)}_{\Omega\vec{k}}\,
  F^*(k_1, \Omega ')\sqrt{\omega_k}
  \left(\frac{\partial}{\partial k_1}
  \sqrt{\omega_k} F(k_1, \Omega)\right) \\
  & &+ (d_p \rightarrow \bar{d}_{\tilde p})\,{.} \nonumber
 \end{eqnarray}
Now we require that the derivative term in (\ref{b4+}) and
(\ref{b4-}) satisfies the relation (see Eq. (\ref{feqdiff}))
 \begin{equation}\label{b5}
   \frac{\partial}{\partial k_1}\left[ \sqrt{\omega_k} \,
   F(k_1, \Omega)\right] =
   i \sigma\,\Omega \, \frac{1}{\sqrt{\omega_k}} \,
   F(k_1, \Omega)\,,
 \end{equation}
where $\sigma=+$ for ${\cal M}_{10}^{(+)}$ and $\sigma=-$ for
 ${\cal M}_{10}^{(-)}$. To find the solutions of (\ref{b5}), we put
 \begin{equation}\label{b7}
   F(k_1, \Omega)=\frac{1}{\sqrt{2\pi\omega_k}}\,
   [g(k_1)]^{i\Omega}\,,
 \end{equation}
so that (\ref{b5}) reduces to the form
 \begin{equation}\label{b8}
   \frac{\partial}{\partial k_1}\, \ln g(k_1)=\frac{\sigma}{\omega_k}\,.
 \end{equation}
A solution is
 \begin{equation}\label{sol1}
   g(k_1)=\left(\frac{\omega_k+k_1}{\omega_k-k_1}\right)^{\sigma/2}\,,
 \end{equation}
from which
  \begin{equation}\label{fexpra1}
  F(\Omega, k_1)=\frac{1}{\sqrt{2\pi\omega_k}}\left(
  \frac{\omega_k+k_1}{\omega_k-k_1}\right)^{i\sigma\,\Omega/2} \,{,}
  \end{equation}
that is Eq. (\ref{fexpr}). We have to bear in mind that the
solutions for $\sigma=+$ and $\sigma=-$ has to be used in the two
distinct sectors, respectively.

Another solution of (\ref{b8}) is
 \begin{equation}\label{sol2}
   {\cal G}(k_1)=e^{i(2n+1)\pi}\,\left(\frac{\omega_k-k_1}{\omega_k+k_1}\right)^{1/2}
   =e^{i(2n+1)}\,[g(k_1)]^{-1}\,,
 \end{equation}
where $n=0, 1, 2, \ldots$.

Using (\ref{b1}) and (\ref{b2}), Eq. (\ref{b4}) can be recast in
the diagonal form
 \begin{equation}\label{b6}
  {\cal M}_{10}=  \sum_{\sigma} \int d\Omega d^{n-2} p\,\sigma\, \Omega
 \,[ d_p^{(\sigma)
 \dagger} d_p^{(\sigma)} + \bar{d}_{\tilde{p}}^{(\sigma)}
 \bar{d}_{\tilde{p}}^{(\sigma)\dagger}]\,{.}
 \end{equation}
Thus, from (\ref{b6}) and using (\ref{17}) and (\ref{18}), Eq.
(\ref{hamrin}) follows.
 \begin{eqnarray}\label{b6bis}
  {\cal H}_\epsilon &=& G(\epsilon){\cal M}_{10}G^{-1}(\epsilon) \\
  &=&  \sum_{\sigma} \int d\Omega d^{n-2} p\,\sigma\, \Omega \,
  [ G(\epsilon)\, d_p^{(\sigma)\dagger}\, G^{-1}(\epsilon)G(\epsilon)
   d_p^{(\sigma)}G^{-1}(\epsilon) \nonumber \\
  & & +
  G(\epsilon)\bar{d}_{\tilde{p}}^{(\sigma)}G^{-1}(\epsilon)
  G(\epsilon)\bar{d}_{\tilde{p}}^{(\sigma)\dagger}G^{-1}(\epsilon)]
 \nonumber \\
  &=&  \sum_{\sigma} \int d\Omega d^{n-2} p\, \sigma\, \Omega \,[
 d_p^{(\sigma)
  \dagger}(\epsilon) d_p^{(\sigma)}(\epsilon) +
  \bar{d}_{\tilde{p}}^{(\sigma)}(\epsilon)
  \bar{d}_{\tilde{p}}^{(\sigma)\dagger}(\epsilon)] \nonumber \,{.}
 \end{eqnarray}
Let us now show that the found solutions for $\{F(k_1, \Omega)\}$
form a complete set. We shall do it for the sector $\sigma=+$. By
putting
 \begin{equation}\label{xdef}
   e^{2x}\equiv \frac{\omega_k+k_1}{\omega_k-k_1}\,,
 \end{equation}
it follows
 \begin{eqnarray}\label{completeset1}
   \int_0^\infty d\Omega F^*(k_1, \Omega)F(k_1',
   \Omega)&=&\frac{1}{2\pi \omega_k}\int_0^\infty d\Omega
   e^{-i\Omega(x-x')}  \\
   &=&\frac{1}{\omega_k}\delta(x-x')=\delta(k_1-k_1') \,. \nonumber
 \end{eqnarray}
In similar way one shows that
  \begin{eqnarray}\label{completeset2}
   \int_{-\infty}^\infty dk_1 F^*(k_1, \Omega ')F(k_1,
   \Omega')&=& \int_0^\infty \frac{dx}{2\pi}\, e^{-ix(\Omega-\Omega ')}
 \\
   &=&\delta(\Omega-\Omega') \,. \nonumber
 \end{eqnarray}

 \end{document}